\documentclass[prd,twocolumn, amsmath,amssymb,floatfix,nofootinbib]{revtex4}
\usepackage{mathrsfs,bm}
\usepackage{longtable,lscape}
\usepackage{txfonts}
\usepackage{indentfirst}
\usepackage{graphicx,color,dcolumn,booktabs}
\usepackage{multirow}
\usepackage{indentfirst}
\usepackage{multirow}
\usepackage{epsfig}
\usepackage{graphicx,color,dcolumn}
\usepackage{epstopdf}
\usepackage{amsmath}
\usepackage{multirow}
\usepackage{diagbox}
\usepackage{changes}
\usepackage{threeparttable}

\usepackage{color}
\usepackage{amssymb}
\definecolor{cover}{rgb}{0.77,0.87,0.88}
\definecolor{blueone}{rgb}{0.1,0.1,.7}
\definecolor{citec}{rgb}{0.14,0.47,0.09}
\definecolor{two}{rgb}{0.0,0.5,0.}
\definecolor{three}{rgb}{.5,.1,0.15}
\usepackage[bookmarks=true,bookmarksopen=false,plainpages=false,breaklinks=true,
   bookmarksnumbered=true,hypertexnames=false,
   filecolor=blue,urlcolor=three,menucolor=three,
   linkcolor=three,citecolor=blueone, colorlinks,
   anchorcolor=blue,runcolor=pink,frenchlinks=red
   pdfstartview=FitH,pdftitle=title,%
   pdfauthor=author]{hyperref}

\usepackage{relsize}
\usepackage{xspace}
\def\babar{\mbox{\slshape B\kern-0.1em{\smaller A}\kern-0.1em
    B\kern-0.1em{\smaller A\kern-0.2em R}}}

\allowdisplaybreaks[4]
\usepackage{color}

\begin{document}

\title{Possible molecular states from  interactions of charmed strange baryons}

\author{Dan Song\footnote{These authors have contributed equally to this
work.\label{aa}}, Shu Chen\footnotemark[1], Shu-Yi Kong and Jun
He\footnote{Corresponding author: junhe@njnu.edu.cn}}

\affiliation{School of Physics and Technology, Nanjing Normal University,
Nanjing 210097, China}

\date{\today}

\begin{abstract}

In this work,  we perform an investigation of possible  molecular states
composed of two charmed strange baryons from the $\Xi_c^{(',*)}{\Xi}_c^{(',*)}$
interaction, and their  hidden-charm hidden-strange partners from the
$\Xi_c^{(',*)}\bar{\Xi}_c^{(',*)}$  interaction. With the help of the heavy
quark chiral effective Lagrangians, the interactions of charmed strange baryons
are described with light meson exchanges. The potential kernels are constructed,
and inserted into the quasipotential Bethe-Salpeter equation. The bound states are produced from most interactions considered,
which suggests that strong attractions exist widely between the charmed strange
baryons. Experimental searching for such molecular states  is suggested in future high-precision measurements.

\end{abstract}

\maketitle

\section{Introduction}

Over the past few decades, more and more  new hadrons were reported in
experiment, which can not be well explained in the frame of the conventional
quark model~\cite{PDG}.  There are a number of explanations for such exotic
hadrons, which include the molecular state picture, compact multiquark picture,
and kinetic mechanisms.  With the experimental observations of more and more
exotic states close to thresholds of two hadrons,  the molecular state picture
is widely  applied to understand the exotic states.

The dibaryons are  important  exotic states, which attract much attention from
hadron physics community since about sixty years ago. The idea of dibaryons was
proposed by Dyson and Xuong theoretically in 1964~\cite{Dyson:1964xwa}. Dibaryon
was really taken seriously after Jaffe's suggestion that there might be
so-called $H$ dibaryon in 1977~\cite{Jaffe:1976yi}. After a long history of ups
and downs in experiment~\cite{Clement:2016vnl}, the resonance $d^*(2380)$ with a
mass of about 2370 MeV and  width of about 70 MeV was observed at
WASA-at-COSY~\cite{WASA-at-COSY:2011bjg}. The resonant structure was studied in
many
pictures~\cite{Gal:2013dca,Huang:2014kja,Haidenbauer:2011za,Park:2015nha,Dong:2015cxa,Dong:2016rva},
including a molecular state from the $\Delta\Delta$
interaction~\cite{Haidenbauer:2011za}. Some authors also suggested that it may
be not a genuine particle, but only a triangle singularity in the last step of
the reaction~\cite{Ikeno:2021frl,Molina:2021bwp,Bar-Nir:1973mxc}. Even if
$d^*(2380)$ is a genuine particle, it is difficult to assign it as a molecular
state because such assumption leads to a binding energy of about 80~MeV, which
tends to assign it as a compact hexaquark rather than a bound state of two
$\Delta$ baryons.  Hence, it is still an interesting topic to find the molecular
state with two baryons beyond the well-known deuteron.  In the past two decades,
many candidates of molecular states in charmed sector were observed,  such as
the hidden-charm pentaquarks
$P_c$~\cite{LHCb:2015yax,LHCb:2019kea,LHCb:2020jpq},  $X(3872)$ and
$Z_c(3900)$~\cite{Belle:2003nnu,Tornqvist:2004qy,BESIII:2013ris,Xiao:2013iha},
which are pretty close to the thresholds of two charmed hadrons. Hence, a
natural direction is to study dibaryon molecular states with heavy quarks,  such
as molecular $N\Omega_{c}$ and $N\Lambda_c$ states proposed in the
literature~\cite{Froemel:2004ea,Liu:2011xc}.

The molecular states were widely studied in the past two decades. However, the
candidates of molecular states composed of two charmed/anticharmed baryons are
still scarcely observed in experiment.  In the literature, there exist some
theoretical works about hidden-charm molecular states composed of a charm and an anticharmed
baryon, especially   $\Lambda_c\bar{\Lambda}_c$  molecular states inspired by
the observation of
$Y(4630)$~\cite{Chen:2011cta,Wang:2021qmn,Lee:2011rka,Simonov:2011jc,Song:2022yfr}.
The double-charm dibaryons also attract much attention from hadron physics
community~\cite{Li:2012bt,Garcilazo:2020acl,Carames:2015sya,Chen:2021cfl,Ling:2021asz,Dong:2021bvy,Song:2022svi}.
In our previous work,  possible molecular states  from  hidden-charm systems
$\Lambda_c\bar{\Lambda}_c$, $\Sigma_c^{(*)}\bar{\Sigma}_c^{(*)}$ and $\Lambda_c
\bar{\Sigma}_c^{(*)}$, and corresponding double-charm systems
$\Lambda_c{\Lambda}_c$, $\Sigma_c^{(*)}{\Sigma}_c^{(*)}$, and $\Lambda_c
{\Sigma}_c^{(*)}$ have already been studied, and strong attractions were found
in these systems~\cite{Song:2022svi}.  In Ref.~\cite{Lee:2011rka}, they drew a
conclusion that there may exist four loosely deuteron-like bound  states
$\Xi_c{\Xi}_c$($\Xi_c\bar{\Xi}_c$) and
$\Xi_c^{'}{\Xi}_c^{'}$($\Xi_c^{'}\bar{\Xi}_c^{'}$). In the current work, we will
study the possible molecular states composed of two charmed strange baryons from
the $\Xi_c^{(',*)}{\Xi}_c^{(',*)}$ interaction, and their  hidden-charm
hidden-strange partners from the $\Xi_c^{(',*)}\bar{\Xi}_c^{(',*)}$  interaction
in a quasipotential Bethe-Salpeter equation (qBSE) approach. 

This article is organized as follows. After introduction, the Lagrangians
depicting the couplings of light mesons and baryons will be presented, and the
potentials are constructed. In addition, the qBSE approach will also be
introduced briefly in Section~\ref{Sec: Formalism}. The numerical results about
the binding energies will be presented in Section~\ref{Sec:Results and
discussions}. Finally, a brief summary are given in Section~\ref{Sec:Summary}.

\section{Theoretical frame}\label{Sec: Formalism}

In the current work, to study the interactions  $\Xi_c^{(',*)}{\Xi}_c^{(',*)}$
and $\Xi_c^{(',*)}\bar{\Xi}_c^{(',*)}$, potential kernel will be constructed
within the one-boson exchange model. The exchanges are intermediated  by
pseudoscalar $\mathbb{P}$ ($\pi$ and $\eta$), vector $\mathbb{V}$ ($\rho$,
$\omega$, and $\phi$), and scalar $\sigma$ mesons. The Lagrangians that depict
the couplings of light mesons and charmed strange baryons are required, and
presented in the below. 

\subsection{Relevant Lagrangians}

The vertices between charmed baryon and light $\pi$, $\rho$,  $\eta$, $\omega$,
$\phi$, $\sigma$  mesons are described by the Lagrangians  with SU(3),  heavy
quark and  chiral symmetry as~\cite{Liu:2011xc,Isola:2003fh,Falk:1992cx},
\begin{align}
{\cal L}_{S}&=-
\frac{3}{2}g_1(v_\kappa)\epsilon^{\mu\nu\lambda\kappa}{\rm tr}[\bar{S}_\mu
{\cal A}_\nu S_\lambda]+i\beta_S{\rm tr}[\bar{S}_\mu v_\alpha (\mathcal{V}^\alpha-
\rho^\alpha)
S^\mu]\nonumber\\
& + \lambda_S{\rm tr}[\bar{S}_\mu F^{\mu\nu}S_\nu]
+\ell_S{\rm tr}[\bar{S}_\mu \sigma S^\mu],\nonumber\\
{\cal L}_{B_{\bar{3}}}&= i\beta_B{\rm tr}[\bar{B}_{\bar{3}}v_\mu(\mathcal{V}^\mu-\rho^\mu)
B_{\bar{3}}]
+\ell_B{\rm tr}[\bar{B}_{\bar{3}}{\sigma} B_{\bar{3}}], \nonumber\\
{\cal L}_{int}&=ig_4 {\rm tr}[\bar{S}^\mu {\cal A}_\mu B_{\bar{3}}]+i\lambda_I \epsilon^{\mu\nu\lambda\kappa}v_\mu{\rm tr}[\bar{S}_\nu F_{\lambda\kappa} B_{\bar{3}}]+h.c.,
\end{align}
where $S^{\mu}_{ab}$ is composed of Dirac spinor operators,
\begin{align}
    S^{ab}_{\mu}&=-\sqrt{\frac{1}{3}}(\gamma_{\mu}+v_{\mu})
    \gamma^{5}B^{ab}+B^{*ab}_{\mu}\equiv{ B}^{ab}_{0\mu}+B^{ab}_{1\mu},\nonumber\\
    \bar{S}^{ab}_{\mu}&=\sqrt{\frac{1}{3}}\bar{B}^{ab}
    \gamma^{5}(\gamma_{\mu}+v_{\mu})+\bar{B}^{*ab}_{\mu}\equiv{\bar{B}}^{ab}_{0\mu}+\bar{B}^{ab}_{1\mu}.
\end{align}
Substituting Dirac spinor operators into the Lagrangians, we reach explicit forms as,
\begin{align}
{\cal L}_{BB\mathbb{P}}&=-i\frac{3g_1}{4f_\pi\sqrt{m_{\bar{B}}m_{B}}}~\epsilon^{\mu\nu\lambda\kappa}\partial^\nu \mathbb{P}~
\sum_{i=0,1}\bar{B}_{i\mu} \overleftrightarrow{\partial}_\kappa B_{j\lambda},\nonumber\\
{\cal L}_{BB\mathbb{V}}&=-\frac{\beta_S g_V}{2\sqrt{2m_{\bar{B}}m_{B}}}\mathbb{V}^\nu
 \sum_{i=0,1}\bar{B}_i^\mu \overleftrightarrow{\partial}_\nu B_{j\mu}\nonumber\\
&-\frac{\lambda_S
g_V}{\sqrt{2}}(\partial_\mu \mathbb{V}_\nu-\partial_\nu \mathbb{V}_\mu) \sum_{i=0,1}\bar{B}_i^\mu B_j^\nu,\nonumber\\
{\cal L}_{BB\sigma}&=\ell_S\sigma\sum_{i=0,1}\bar{B}_i^\mu B_{j\mu},\nonumber\\
    {\cal L}_{B_{\bar{3}}B_{\bar{3}}\mathbb{V}}&=-\frac{g_V\beta_B}{2\sqrt{2m_{\bar{B}_{\bar{3}}}m_{B_{\bar{3}}}} }\mathbb{V}^\mu\bar{B}_{\bar{3}}\overleftrightarrow{\partial}_\mu B_{\bar{3}},\nonumber\\
{\cal L}_{B_{\bar{3}}B_{\bar{3}}\sigma}&=i\ell_B \sigma \bar{B}_{\bar{3}}B_{\bar{3}},\nonumber\\
{\cal L}_{BB_{\bar{3}}\mathbb{P}}
    &=-i\frac{g_4}{f_\pi} \sum_i\bar{B}_i^\mu \partial_\mu \mathbb{P} B_{\bar{3}}+{\rm H.c.},\nonumber\\
{\cal L}_{BB_{\bar{3}}\mathbb{V}}    &=\frac{g_\mathbb{V}\lambda_I}{\sqrt{2m_{\bar{B}}m_{B_{\bar{3}}}}} \epsilon^{\mu\nu\lambda\kappa} \partial_\lambda \mathbb{V}_\kappa\sum_i\bar{B}_{i\nu} \overleftrightarrow{\partial}_\mu
   B_{\bar{3}}+{\rm H.c.}.
   \label{LB}
\end{align}

The charmed baryon matrices are defined as
\begin{align}
B_{\bar{3}}&=\left(\begin{array}{ccc}
0&\Lambda^+_c&\Xi_c^+\\
-\Lambda_c^+&0&\Xi_c^0\\
-\Xi^+_c&-\Xi_c^0&0
\end{array}\right),\quad
B=\left(\begin{array}{ccc}
\Sigma_c^{++}&\frac{1}{\sqrt{2}}\Sigma^+_c&\frac{1}{\sqrt{2}}\Xi'^+_c\\
\frac{1}{\sqrt{2}}\Sigma^+_c&\Sigma_c^0&\frac{1}{\sqrt{2}}\Xi'^0_c\\
\frac{1}{\sqrt{2}}\Xi'^+_c&\frac{1}{\sqrt{2}}\Xi'^0_c&\Omega^0_c
\end{array}\right). \nonumber\\
B^*&=\left(\begin{array}{ccc}
\Sigma_c^{*++}&\frac{1}{\sqrt{2}}\Sigma^{*+}_c&\frac{1}{\sqrt{2}}\Xi^{*+}_c\\
\frac{1}{\sqrt{2}}\Sigma^{*+}_c&\Sigma_c^{*0}&\frac{1}{\sqrt{2}}\Xi^{*0}_c\\
\frac{1}{\sqrt{2}}\Xi^{*+}_c&\frac{1}{\sqrt{2}}\Xi^{*0}_c&\Omega^{*0}_c
\end{array}\right).\label{MBB}
\end{align}

The $\mathbb{V}$ and $\mathbb{P}$ are the vector and pseudoscalar matrices as
\begin{align}
{\mathbb P}=\left(\begin{array}{ccc}
        \frac{\sqrt{3}\pi^0+\eta}{\sqrt{6}}&\pi^+&K^+\\
        \pi^-&\frac{-\sqrt{3}\pi^0+\eta}{\sqrt{6}}&K^0\\
        K^-&\bar{K}^0&\frac{-2\eta}{\sqrt{6}}
\end{array}\right),
\mathbb{V}=\left(\begin{array}{ccc}
\frac{\rho^{0}+\omega}{\sqrt{2}}&\rho^{+}&K^{*+}\\
\rho^{-}&\frac{-\rho^{0}+\omega}{\sqrt{2}}&K^{*0}\\
K^{*-}&\bar{K}^{*0}&\phi
\end{array}\right).\nonumber
\end{align}

We choose the central values suggested in the Review of  Particle Physics  (PDG)~\cite{PDG} as the masses of  particles involved in the calculation. The values for different charges will be averaged. The  coupling constants involved are listed in Table~\ref{coupling}. 
\renewcommand\tabcolsep{0.12cm}
\renewcommand{\arraystretch}{1.}
\begin{table}[h!]
\caption{The coupling constants adopted in the
calculation, which are cited from the literature~\cite{Chen:2019asm,Liu:2011xc,Isola:2003fh,Falk:1992cx,Zhu:2022fyb,Zhu:2021lhd}. The $\lambda$ and $\lambda_{S,I}$ are in the units of GeV$^{-1}$. Others are in the units of $1$.
\label{coupling}}
\begin{tabular}{cccccccccccccccccc}\bottomrule[1pt]
$\beta$&$g$&$g_V$&$\lambda$ &$g_{s}$\\
0.9&0.59&5.9&0.56 &0.76\\\hline
$\beta_S$&$\ell_S$&$g_1$&$\lambda_S$ &$\beta_B$&$\ell_B$ &$g_4$&$\lambda_I$\\
-1.74&6.2&-0.94&-3.31&$-\beta_S/2$&$-\ell_S/2$&$3g_1/{(2\sqrt{2})}$&$-\lambda_S/\sqrt{8}$ \\
\bottomrule[1pt]
\end{tabular}
\end{table}

The flavor wave functions for the double-charm double-strange system $\Xi_c^{(',*)}{\Xi}_c^{(',*)}$ can be written as,
\begin{align}
|\Xi_c^{(',*)}{\Xi}_c^{(',*)}\rangle|_{I=0}&=\frac{1}{\sqrt{2}}\left(|\Xi_c^{(',*)+}{\Xi}_c^{(',*)0}\rangle-|\Xi_c^{(',*)0}{\Xi}_c^{(',*)+}\rangle\right),\nonumber\\
|\Xi_c^{(',*)}{\Xi}_c^{(',*)}\rangle|_{I=1}&=|\Xi_c^{(',*)+}{\Xi}_c^{(',*)+}\rangle.
\end{align}

Following the method in Ref.~\cite{He:2019rva}, we input vertices $\Gamma$ and  propagators $P$  into the code directly.  The  explicit forms of potential  can be written with the Lagrangians and flavor wave functions as,
\begin{equation}%
{\cal V}_{\mathbb{P},\sigma}=f_I\Gamma_1\Gamma_2 P_{\mathbb{P},\sigma}f(q^2),\ \
{\cal V}_{\mathbb{V}}=f_I\Gamma_{1\mu}\Gamma_{2\nu}  P^{\mu\nu}_{\mathbb{V}}f(q^2).\label{V}
\end{equation}
The propagators are defined as usual as
\begin{equation}%
P_{\mathbb{P},\sigma}= \frac{i}{q^2-m_{\mathbb{P},\sigma}^2},\ \
P^{\mu\nu}_\mathbb{V}=i\frac{-g^{\mu\nu}+q^\mu q^\nu/m^2_{\mathbb{V}}}{q^2-m_\mathbb{V}^2},
\end{equation}
where $f(q^2)$ is a form factor  adopted to compensate the off-shell effect of
exchanged meson. The form factor is shown as
$f(q^2)=e^{-(m_e^2-q^2)^2/\Lambda_e^2}$, where the $m_e$ and $q$ are the mass
and momentum of the exchanged  meson, respectively.  The $f_I$ is the flavor
factor for certain meson exchange of certain interaction, and the explicit
values are listed in Table~\ref{flavor factor}.

The potential kernels of a charmed strange and an anticharmed antistrange baryon
are  different from the double-charm double-strange systems. However, the former
potential kernel $\bar{\cal V}$ can be written from the latter one ${\cal V}$, which has been calculated above, with the
help of  well-known $G$-parity rule.  However, before applying such treatment, the $C$ or $G$ parity should be introduced for the system $\Xi_c^{(')}\bar{\Xi}_c^{*}$. We take the
following charge conjugation convention as in Ref.~\cite{Dong:2021juy},
\begin{equation}
|B_1B_2\rangle_c=\frac{1}{\sqrt{2}}|B_1\bar{B}_2\rangle-(-1)^{J-J_1-J_2}cc_1c_2|B_2\bar{B}_1\rangle,
\end{equation}
where $J_1$ and $J_2$ are the spins of baryons $|B_1\rangle$ and $|B_2\rangle$,
respectively. $J$ is the total angular momentum of the coupling of $J_1$ and
$J_2$, $c_i$ is defined by $\mathcal{C}|B_i\rangle = c_i\bar{B}_i\rangle$. For
the isovector states, the $C$ parity can not be defined, so we will use the $G$
parity  instead, as $G=(-1)^{I}C$. 
By inserting the $G^{-1}G$ operator into
the potential, the $G$-parity rule can be obtained easily
as~\cite{Phillips:1967zza,Klempt:2002ap,Lee:2011rka,Zhu:2019ibc},
\begin{eqnarray} \bar{\cal V}&=&\sum_{i}{\zeta_{i}{\cal V}_{i}}.  \end{eqnarray}
The $G$ parity of the exchanged meson is left as a $\zeta_{i}$ factor for meson
$i$.

\renewcommand\tabcolsep{0.11cm}
\renewcommand{\arraystretch}{1.5}
\begin{table}[h!]
\begin{center}
\caption{The isospin factors $I_i^d$  and $(-1)^{(J+1)} I_i^c$ of exchange $i$ for direct diagrams and cross diagrams, respectively. The values in bracket are for the case double-charm baryons.}
\label{flavor factor}
\begin{tabular}{c|c|cccccc}\bottomrule[1pt]
  $I_i^d$ &$I$&$\pi$&$\eta$  &$\rho$ &$\omega$&$\phi$&$\sigma$  \\\hline
$\Xi_c \bar{\Xi}_c$&$0$&$ $&$ $&$\frac{3}{2}[-\frac{3}{2}]$ &$-\frac{1}{2}[\frac{1}{2}]$ &$-1[1]$&$4[4]$\\
$ $&$1$&$ $&$ $&$-\frac{1}{2}[\frac{1}{2}]$&$-\frac{1}{2}[\frac{1}{2}]$&$-1[1]$&$4[4]$\\
$\Xi_c^{(',*)}  \bar{\Xi}_c^{(',*)}$&$0$&$-\frac{3}{8}[-\frac{3}{8}]$&$\frac{1}{24}[\frac{1}{24}]$&$\frac{3}{8}[-\frac{3}{8}]$&$-\frac{1}{8}[\frac{1}{8}]$&$-\frac{1}{4}[\frac{1}{4}]$& $1[1]$\\
$ $&$1$&$\frac{1}{8}[\frac{1}{8}]$&$\frac{1}{24}[\frac{1}{24}]$&$-\frac{1}{8}[\frac{1}{8}]$&$-\frac{1}{8}[\frac{1}{8}]$&$-\frac{1}{4}[\frac{1}{4}]$& $1[1]$\\
$\Xi_c \bar{\Xi}_c^{(',*)}$&$0$&$ $&$ $&$\frac{3}{4}[-\frac{3}{4}]$&$-\frac{1}{4}[\frac{1}{4}]$&$-\frac{1}{2}[\frac{1}{2}]$&$2[2]$\\
$ $&$1$&$ $&$ $&$-\frac{1}{4}[\frac{1}{4}]$&$-\frac{1}{4}[\frac{1}{4}]$&$-\frac{1}{2}[\frac{1}{2}]$&$2[2]$\\
\bottomrule[1pt]
  $(-1)^{(J+1)} I_i^c$  &$I$&$\pi$&$\eta$  &$\rho$ &$\omega$&$\phi$ \\\hline
$\Xi_c \bar{\Xi}_c^{'}$&$0$&$\frac{3c}{4}[-\frac{3}{4}]$&$-\frac{3c}{4}[\frac{3}{4}]$&$-\frac{3c}{4}[-\frac{3}{4}]$&$\frac{c}{4}[\frac{1}{4}]$&$\frac{c}{2}[\frac{1}{2}]$  \\
$ $&$1$&$\frac{c}{4}[\frac{1}{4}]$&$\frac{3c}{4}[\frac{3}{4}]$&$-\frac{c}{4}[\frac{1}{4}]$&$-\frac{c}{4}[\frac{1}{4}]$&$-\frac{c}{2}[\frac{1}{2}]$\\
$\Xi_c \bar{\Xi}_c^*$&$0$&$-\frac{3c}{4}[-\frac{3}{4}]$&$\frac{3c}{4}[\frac{3}{4}] $&$\frac{3c}{4}[-\frac{3}{4}]$ &$-\frac{c}{4}[\frac{1}{4}] $&$-\frac{c}{2}[\frac{1}{2}]$ \\
$ $&$1$&$\frac{c}{4}[\frac{1}{4}]$&$\frac{3c}{4}[\frac{3}{4}]$&$-\frac{c}{4}[\frac{1}{4}]$&$-\frac{c}{4}[\frac{1}{4}]$&$-\frac{c}{2}[\frac{1}{2}]$\\
$\Xi_c^{'}  \bar{\Xi}_c^*$&$0$&$-\frac{3c}{8}[-\frac{3}{8}]$&$\frac{c}{24}[\frac{1}{24}]$&$\frac{3c}{8}[-\frac{3}{8}]$&$-\frac{c}{8}[\frac{1}{8}]$&$-\frac{c}{4}[\frac{1}{4}]$ \\
$ $&$1$&$-\frac{c}{8}[\frac{1}{8}] $&$-\frac{c}{24}[\frac{1}{24}] $&$\frac{c}{8}[\frac{1}{8}] $ &$\frac{c}{8}[-\frac{1}{8}] $&$-\frac{c}{4}[\frac{1}{4}]$  \\
\toprule[1pt]
\end{tabular}
\end{center}
\end{table}

To solve the scattering amplitude, the potential kernel obtained above is
adopted to solve the
qBSE~\cite{He:2014nya,He:2015mja,He:2012zd,He:2015yva,He:2017aps,Zhu:2021lhd,Kong:2021ohg,Ding:2020dio}.
The 4-dimensional integral equation in the Minkowski space can be reduced to a
1-dimensional integral equation with fixed spin-parity $J^P$
as~\cite{He:2015mja} by partial-wave decomposition and spectator quasipotential
approximation. The 1-dimensional Bethe-Saltpeter equation is further converted
to  a matrix equation, the scattering amplitude can be obtained as~\cite{He:2014nya,He:2015mja,He:2012zd,He:2015yva,He:2017aps,Zhu:2021lhd,Kong:2021ohg,Ding:2020dio},
\begin{align}
i{\cal M}^{J^P}_{\lambda'\lambda}({\rm p}',{\rm p})
&=i{\cal V}^{J^P}_{\lambda',\lambda}({\rm p}',{\rm
p})+\sum_{\lambda''}\int\frac{{\rm
p}''^2d{\rm p}''}{(2\pi)^3}\nonumber\\
&\cdot
i{\cal V}^{J^P}_{\lambda'\lambda''}({\rm p}',{\rm p}'')
G_0({\rm p}'')i{\cal M}^{J^P}_{\lambda''\lambda}({\rm p}'',{\rm
p}),\quad\quad \label{Eq: BS_PWA}
\end{align}
where the sum extends only over nonnegative helicity $\lambda''$.
The $G_0({\rm p}'')$ is reduced from the 4-dimensional  propagator under quasipotential approximation as  $G_0({\rm p}'')=\delta^+(p''^{~2}_h-m_h^{2})/(p''^{~2}_l-m_l^{2})$ with $p''_{h,l}$ and $m_{h,l}$ being the momentum and mass of heavy or light constituent particle.
The partial wave potentials are obtained with the potentials obtained above in Eq.~(\ref{V}) as
\begin{align}
{\cal V}_{\lambda'\lambda}^{J^P}({\rm p}',{\rm p})
&=2\pi\int d\cos\theta
~[d^{J}_{\lambda\lambda'}(\theta)
{\cal V}_{\lambda'\lambda}({\bm p}',{\bm p})\nonumber\\
&+\eta d^{J}_{-\lambda\lambda'}(\theta)
{\cal V}_{\lambda'-\lambda}({\bm p}',{\bm p})],
\end{align}
where $\eta=PP_1P_2(-1)^{J-J_1-J_2}$ with the $P$ and $J$ being parity and spin for system, respectively. The initial relative momentum is chosen as ${\bm p}=(0,0,{\rm p})$  and the final one as ${\bm p}'=({\rm p}'\sin\theta,0,{\rm p}'\cos\theta)$. The $d^J_{\lambda\lambda'}(\theta)$ is the Wigner $d$-matrix.
We also adopt an  exponential regularization  by introducing a form factor into the propagator as
$G_0({\rm p}'')\to G_0({\rm p}'')\left[e^{-(p''^2_l-m_l^2)^2/\Lambda_r^4}\right]^2$ with $\Lambda_r$ being a cutoff~\cite{He:2015mja}.

\section{Numerical Results}\label{Sec:Results and discussions}

With previous preparation, the numerical calculation can be performed on the
systems mentioned above, which is to study the molecular states from the
interactions $\Xi_c^{(',*)}\bar{\Xi}_c^{(',*)}$ and $\Xi_c^{(',*)}\Xi_c^{(',*)}$.  The
molecular states can be searched for as the poles of the scattering amplitude in
the complex energy plane. The parameters of the Lagrangians in the current work
are chosen as the same as those in our previous study of the hidden-charm
pentaquarks~\cite{He:2019rva,He:2019ify}, and has been listed in
Table~\ref{coupling}.  The only free parameters are cutoff $\Lambda_e$ and
$\Lambda_r$, which are rewritten as a form of
$\Lambda_e=\Lambda_r=m_e+\alpha~0.22$~GeV with the $m_e$ being the mass of exchanged meson between two constituent baryons. Hence, in the current work, only one
free parameter $\alpha$ is involved in explicit calculation, which is
constrained in a reasonable range from  0 to 2. In the current work, only S-wave
bound states will be considered.

\subsection{Interactions $\Xi_c^{(')} \bar{\Xi}^{(')}_c$ and $\Xi^{(')}_c\Xi^{(')}_c$  }

Since isospin and spin of the $\Xi_c$ baryon are the same as those of the $\Xi'_c$
baryon, the same number of channels will be involved in  interactions
$\Xi_c^{(')} \bar{\Xi}^{(')}_c$ and $\Xi^{(')}_c\Xi^{(')}_c$.  The $\Xi_c^{(')}$
baryon carries spin 1/2 and isospin 1/2,  hence, the total angular momentum of
the systems can be  $J$=0 or $1$ in S-wave channel, and the total isospin can be
0 and 1.  According to the Pauli exclusion principle, the isoscalar axial-vector $\Xi^{(')}_c\Xi^{(')}_c$ state, as well as  isovector scalar $\Xi^{(')}_c\Xi^{(')}_c$ state, does not exist.

\begin{figure}[htpb!]
  \centering
  \includegraphics[scale=0.6,bb=81 200 500 530,clip]{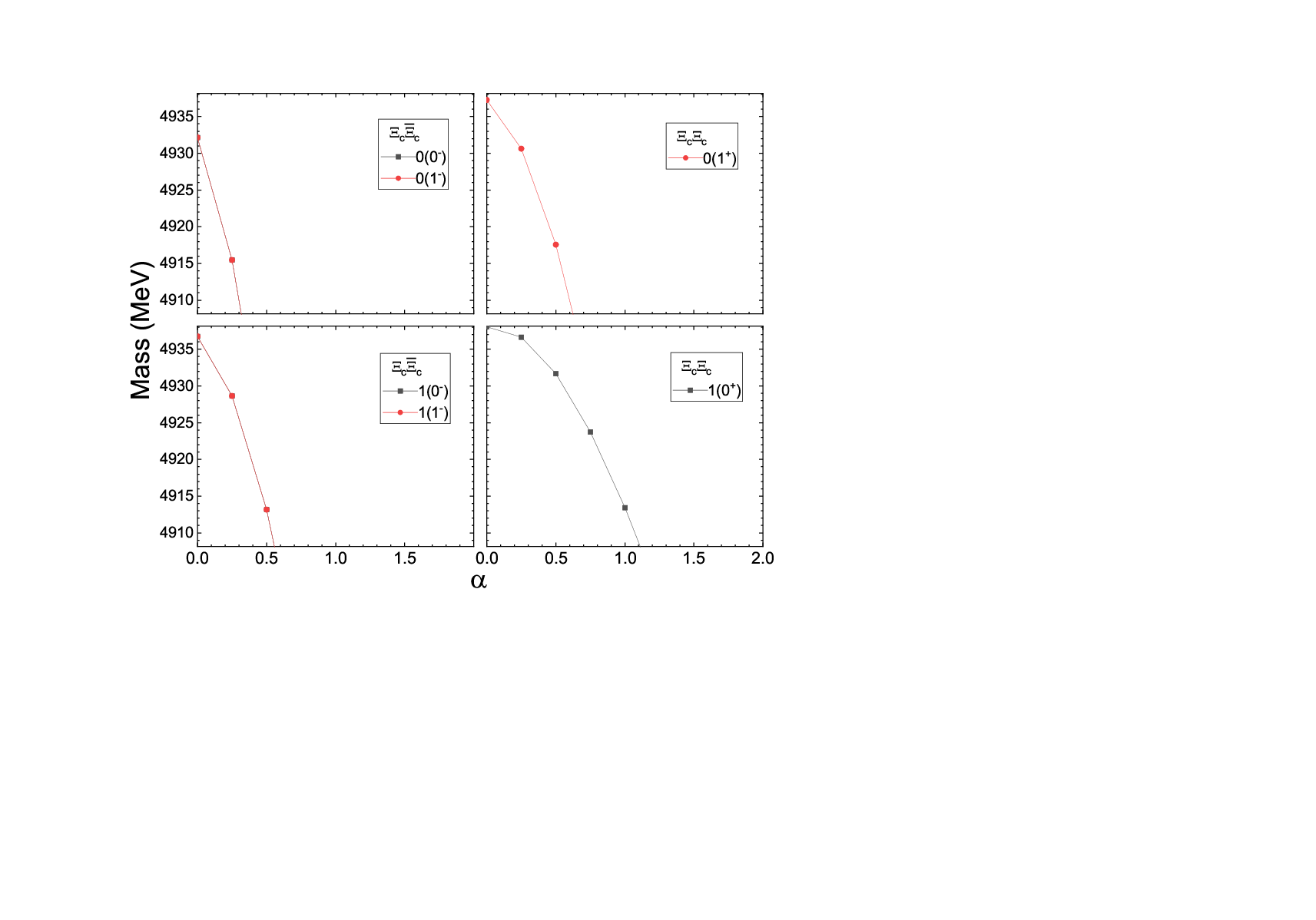}\\
  \caption{The  masses of the bound states from the  interactions $\Xi_c \bar{\Xi}_c$ (left) and $\Xi_c\Xi_c$ (right) with threshold of 4938.15~MeV with the variation of parameter $\alpha$.}\label{0}
\end{figure}

For the interactions $\Xi_c \bar{\Xi}_c$ and $\Xi_c \Xi_c$ considered, the
masses of produced  bound states with the variation of parameter
$\alpha$ are shown in Fig.~\ref{0}. 
The results suggest that isoscalar bound states are produced from the  $\Xi_c
\bar{\Xi}_c$ interaction with total angular momenta $J$=1 and 0. The masses of these two states are almost the same at the same $\alpha$ values. The states appear even at $\alpha$ values below 0, and the masses decrease rapidly with the increase of $\alpha$ value. It suggests a
strong attraction existing between two baryons.  The bound state can be also
found from the double-charm system. The bound state appears at $\alpha$
values about 0, and corresponding mass decreases slower than those of 
hidden-charm sates.  For the isovector interactions, the bound states appear at
$\alpha$ values about 0 for both hidden-charm and double-charm interactions. Their
masses show a slower decrease compared with the isoscalar
interactions.

\begin{figure}[htpb!]
  \centering
  \includegraphics[scale=0.6,bb=78 209 500 530,clip]{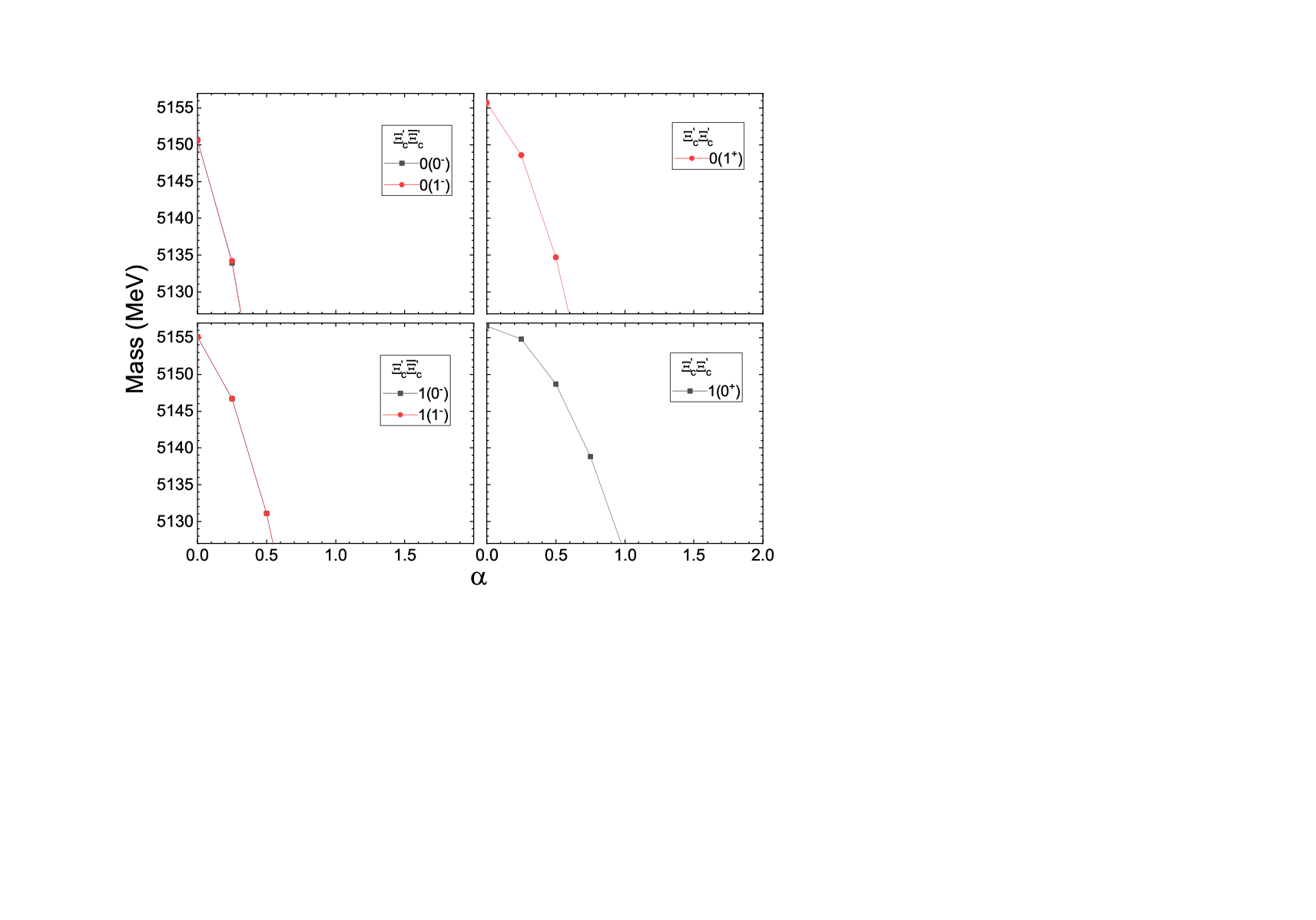}\\
  \caption{The  masses of the bound states from the  interactions $\Xi'_c \bar{\Xi}'_c$ (left) and $\Xi'_c\Xi'_c$ (right) with threshold of 5156.9~MeV with the variation of parameter $\alpha$. }\label{scsc}
\end{figure}

In Fig.~\ref{scsc}, the bound states from the $\Xi_c^{'} \bar{\Xi}_c^{'}$
interaction and their double-charm partners are presented. Since the flavor
factors are the same as those for the interactions  $\Xi_c \bar{\Xi}_c$ and
$\Xi_c\Xi_c$,  bound states are also produced in all channels, and  the masses exhibit similar behaviors as the results in Fig.~\ref{0}. For the
isoscalar hidden-charm $\Xi_c^{'}\bar{\Xi}_c^{'}$ system, the bound states are
produced at  $\alpha$ values below 0, which are smaller than those for the
double-charm $\Xi_c^{'}\Xi_c^{'}$ interaction, which indicts that the
$\Xi_c^{'}\bar{\Xi}_c^{'}$ interaction is more attractive than the
$\Xi_c^{'}\Xi_c^{'}$ interaction due to different contributions from the meson
exchanges. As in the case of $\Xi_c\bar{\Xi}_c$ interaction, the masses for corresponding hidden-charm systems increas faster than the
double-charm system. 

\begin{figure}[htpb!] \centering
\includegraphics[scale=0.6,bb=80 210 500 530,clip]{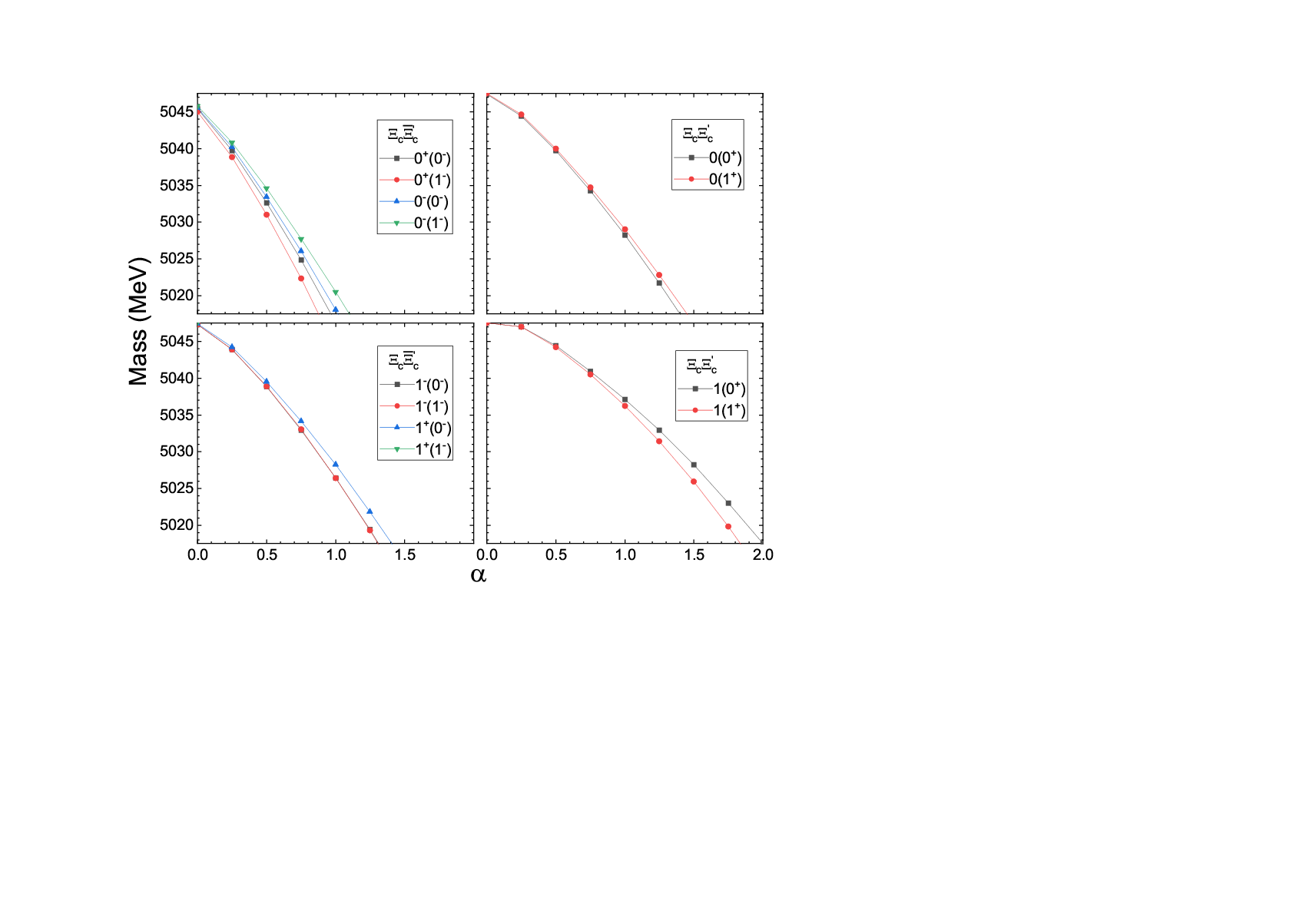}\\ \caption{The
masses of the bound states from the   interactions $\Xi_c\bar{\Xi}_c^{'}$ (left) and
$\Xi_c\Xi_c^{'}$ (right) with threshold of 5047.525~MeV with the variation of parameter $\alpha$. 
\label{xcxcp}} \end{figure}

The masses of the states produced from the $\Xi_c\bar{\Xi}_c'$
interaction and their double-charm partners are shown in Fig.~\ref{xcxcp}. Since
the two constituent baryons are different,  the $C$ parity  involves in the
hidden-charm sector, while it does not involve in the double-charm sector.  For
the hidden-charm system with $I$=0, the bound states are produced at $\alpha$
values below 0. The corresponding double-charm partners appear at  larger
$\alpha$ values.  The hidden-charm bound states with $I$=1 appear at $\alpha$
values below 0 while their double-charm partner appears at $\alpha$ values about
0 and their masses increas much slowly than those of hidden-charm
states.

\subsection{Interactions $\Xi_c^{(')}\bar{\Xi}_c^{*}$ and $\Xi_c^{(')}\Xi_c^{*}$  }

\begin{figure}[htpb!] \centering
\includegraphics[scale=0.6,bb=80 205 500 530,clip]{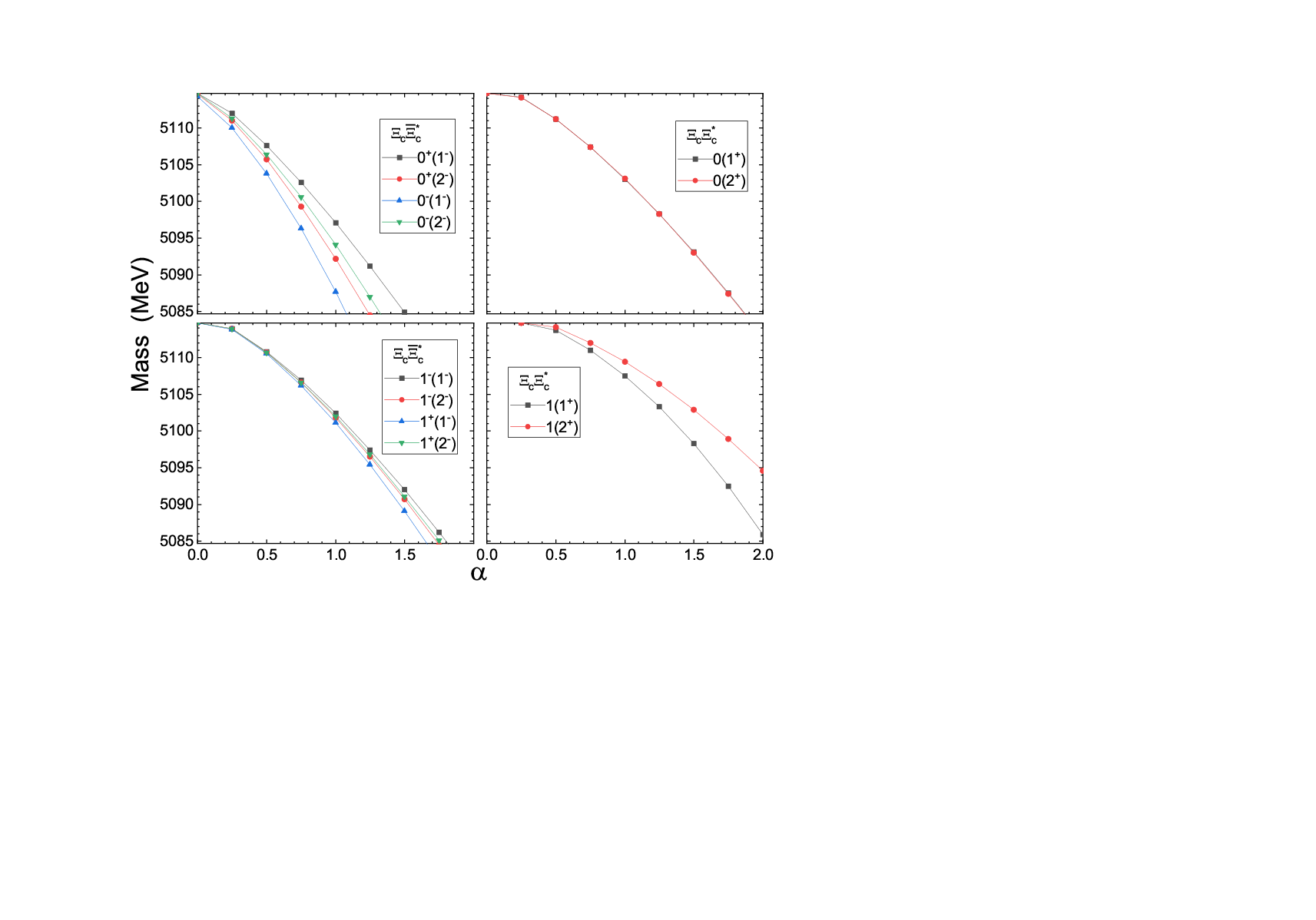}\\ \caption{The
masses of the bound states from   interactions $\Xi_c\bar{\Xi}_c^*$ (left) and
$\Xi_c\Xi_c^*$ (right) with threshold of 5114.7~MeV  with the variation of parameter $\alpha$. 
}\label{lcsc} \end{figure}

Now, we turn to the systems with a  $\Xi/\Xi'$ and a  $\Xi^{*}$ baryon. Since
these baryons carry spins of 1/2 and 3/2, respectively,  the S-wave states have
total angular momenta $J$ = 1 and 2, and will be considered in the calculation. 
The isospins of the $\Xi_c^{(')}\bar{\Xi}_c^{*}$ and $\Xi_c^{(')}\Xi_c^{*}$ systems are the same as  those of the $\Xi_c^{(')} \bar{\Xi}^{(')}_c$ and
$\Xi^{(')}_c\Xi^{(')}_c$ systems.  In these systems, two constituent baryons are
different, hence, the $C$ parity will involve in the hidden-charm sector. The
masses of the $\Xi_c\bar{\Xi}_c^{(*)}$ states and their double-charm
partners are shown in Fig.~\ref{lcsc}.  
The hidden-charm states with $I$=0 appears at 
$\alpha$ values about 0, and the binding energies increase to 30~GeV at 
$\alpha$ values about 1.25. Their double-charm partners also appear at 
$\alpha$ values about 0, but reach a binding energy bout 30~MeV at  larger
$\alpha$ values, about 2. And two double-charm states with $J$=1 and 0 have
almost the same mass. The isovector states also appear at  $\alpha$
values about 0. The double-charm states appear at  $\alpha$ values a little
larger and the binding energies increase slowly, reach 30 MeV at $\alpha$
values larger than 2 as in the isoscalar case.

\begin{figure}[htpb!]
  \centering
  \includegraphics[scale=0.6,bb=80 210 500 530,clip]{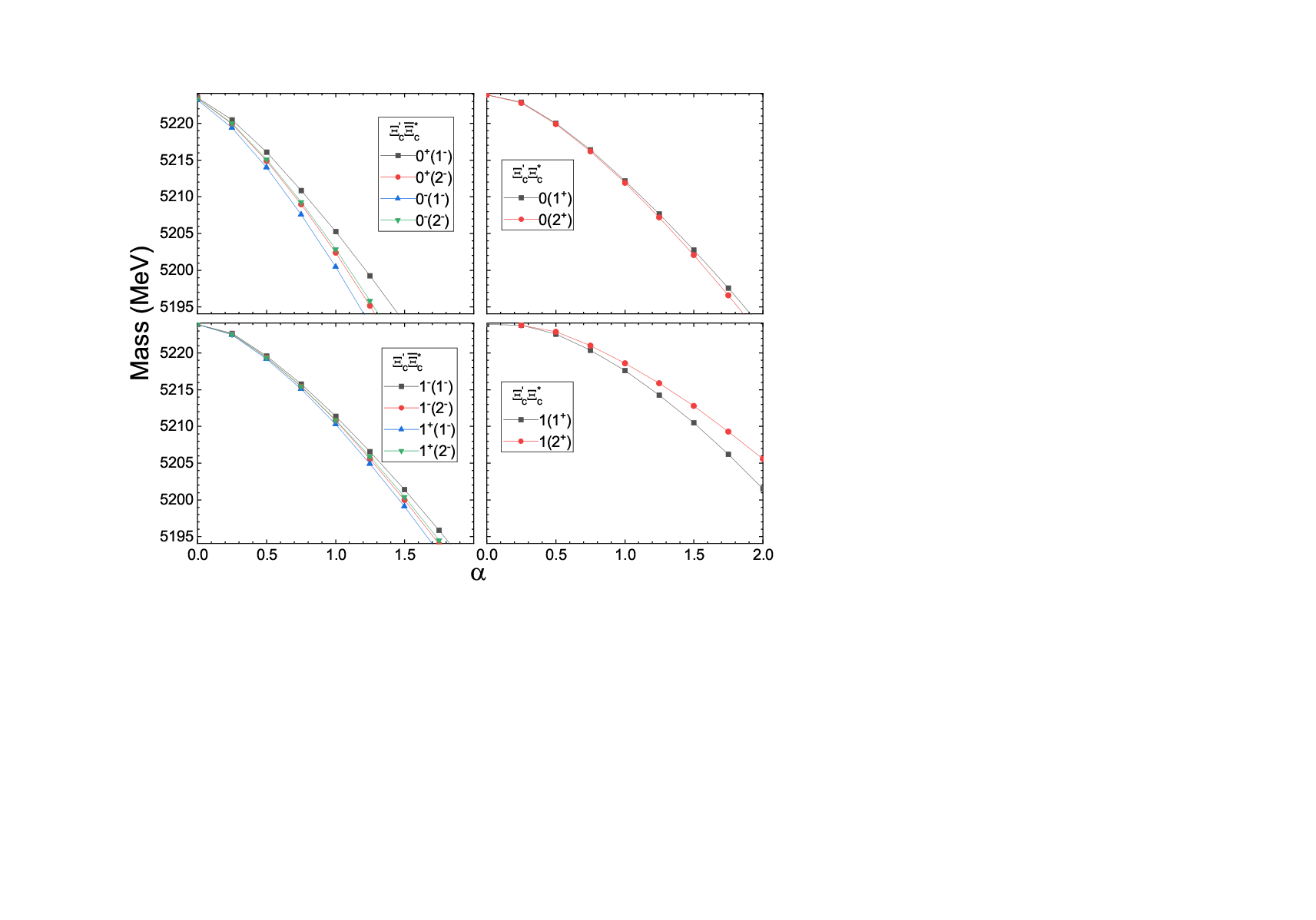}\\
  \caption{The  mases of the bound states from the  interactions $\Xi_c^{'} \bar{\Xi}_c^*$ (left) and $\Xi_c^{'}\Xi_c^*$ (right) with threshold of 5224.08~MeV with the variation of parameter $\alpha$.  }\label{scsca}
\end{figure}

The results for the  interactions $\Xi'_c \bar{\Xi}_c^*$ and $\Xi'_c\Xi_c^*$ are
presented in Fig~\ref{scsca}.  The $C$ parity also involves in the hidden-charm
systems. For the hidden-charm systems with $I=0$, the bound states appear at 
$\alpha$ values about 0 and the binding energies increase to 30 MeV at $\alpha$
values  about 1.25. Their double-charm partners appear at  $\alpha$ values
larger than 0, and the binding energies increase to 30 MeV at $\alpha$ values
about 1.9. In the isovector case, the states appear at $\alpha$ values about 0,
and the binding energies increase to 30 MeV at $\alpha$ values about 1.75. For
their double-charm partners, the bound states appear at $\alpha$ values larger
than 0, and the binding energies increase more slowly than those of hidden charm
states.

\subsection{Interactions $\Xi_c^* \bar{\Xi}_c^{*}$ and $\Xi^*_c \Xi_c^{*}$  }

\begin{figure}[htpb!]
  \centering
  \includegraphics[scale=0.6,bb=80 210 500 530,clip]{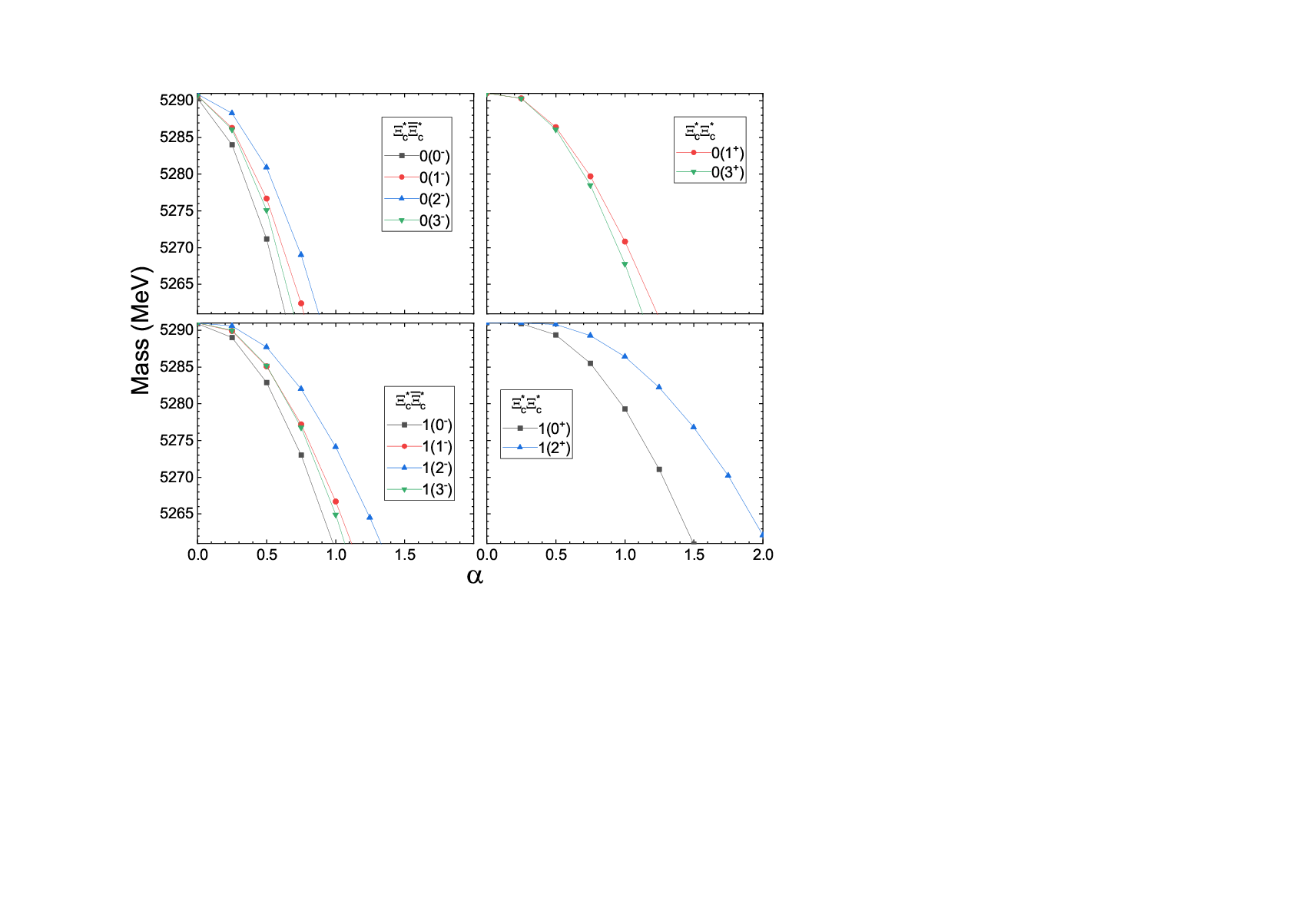}\\
  \caption{The masses of the bound states from the  interactions $\Xi_c^* \bar{\Xi}_c^*$ (left) and $\Xi_c^* \Xi_c^* $ (right) with threshold of 5291.26~MeV with the variation of parameter $\alpha$. }\label{scasca}
\end{figure}

The masses of the states produced from the interactions $\Xi_c^*
\bar{\Xi}_c^*$  and $\Xi^*_c \Xi_c^{*}$  are presented in Fig.~\ref{scasca}.
Since both baryons carry spin 3/2, there are four total angular momenta $J$=0,
1, 2, and 3.  According to the Pauli exclusion principle, the isoscalar $\Xi^{(')}_c\Xi^{(')}_c$ states with $J=$2, 4, as well as  isovector  $\Xi^{(')}_c\Xi^{(')}_c$ states with $J=$1, 3, do not exist.   For the hidden-charm system with $I$=0, there are four states
produced at $\alpha$ values about 0. For the double-charm system with
$I$=0, the binding energies increase to 30 MeV at $\alpha$ values about 1.25.
The hidden-charm states with $I$=1 appear at $\alpha$ values about 0 while their
double-charm partners appear at $\alpha$ values about 0.25 or larger. For the
hidden-charm system with $I$=1, the binding energies increase to 30 MeV at
$\alpha$ values about 0.75, but for the $\Xi_c^*\Xi_c^*$ states, their binding
energies increase to 30 MeV at $\alpha$ values about 1.75 or larger.

\section{Summary}\label{Sec:Summary}

In the current work, we systematically study the molecular states from
interactions of charmed strange baryons, which include  hidden-charm
hidden-strange systems $\Xi_c^{(',*)}\bar{\Xi}_c^{(',*)}$ as well as their
double-charm double-strange partners $\Xi_c^{(',*)}\Xi_c^{(',*)}$. The
calculation indicates that  attractions exist widely in systems composed
of two charmed strange baryons.  For the $\Xi_c\bar{\Xi}_c$ interaction, the
bound states are produced at $\alpha$ values about 0, and the binding energies of their double-charm
partner increase to 30~MeV at  $\alpha$ values a litter larger.  The situations
for interactions $\Xi_c^{'}\Xi_c^{'}$ and $\Xi_c\Xi_c$ are quite similar and the
attractions in the hidden-charm systems $\Xi_c^{'}\bar{\Xi}_c^{'}$ are
stronger than those in the double-charm systems.  For the states with  the same
angular momenta, the masses for states with different isospins behave
almost the same as the variations of parameter $\alpha$. However, for
interactions $\Xi_c^{*}\Xi_c^{*}$ and $\Xi_c^{'}\Xi_c^{*}$, the masses
of states with different isospins exhibit different behaviors with the variation
of parameter.

Generally speaking, many interactions of the two charmed strange baryons are
found attractive, and many bound states are
predicted~\cite{Chen:2011cta,Wang:2021qmn,Lee:2011rka,Simonov:2011jc,Song:2022yfr,Song:2022svi}.
However, up to date, the candidates of the hidden-charm molecular states
composed of two baryons  are scarcely observed experimentally, and no candidates
of the double-charm  molecular states composed of two baryons is reported. It is
quite easy to understand because two baryons, which have six quarks including
two heavy quarks, should be produced simultaneously with a small relative
momentum to form a bound state, which makes the experimental production of such
states more difficult than these composed of two mesons.  However, with the
development of the experimental techniques and accumulation of data at
facilities, such as LHCb,  observation of these states may be expected in the
future. Hence, experimental searching for such molecular states  is suggested in the
future high-precision experiments.

\noindent {\bf Acknowledgement} This project is supported by the National Natural Science
Foundation of China (Grants No. 11675228).


\begin{thebibliography}{99}%
  %1\cite{Tanabashi:2018oca}
\bibitem{PDG}
R.L. Workman {\it et al.} [Particle Data Group],
  ``Review of Particle Physics,''
  Prog. Theor. Exp. Phys. {\bf 2022}, 083C01 (2022)
  %doi:10.1103/PhysRevD.98.030001
  %%CITATION = doi:10.1103/PhysRevD.98.030001;%%
  %2674 citations counted in INSPIRE as of 12 Oct 2019

%\cite{Dyson:1964xwa}
\bibitem{Dyson:1964xwa}
F.~Dyson and N.~H.~Xuong,
``Y=2 States in Su(6) Theory,''
Phys. Rev. Lett. \textbf{13}, no.26, 815-817 (1964)
%doi:10.1103/PhysRevLett.13.815
%190 citations counted in INSPIRE as of 06 Apr 2023
%\cite{Jaffe:1976yi}
\bibitem{Jaffe:1976yi}
R.~L.~Jaffe,
``Perhaps a Stable Dihyperon,''
Phys. Rev. Lett. \textbf{38}, 195-198 (1977)
[erratum: Phys. Rev. Lett. \textbf{38}, 617 (1977)]
%doi:10.1103/PhysRevLett.38.195
%1356 citations counted in INSPIRE as of 06 Apr 2023

%\cite{Clement:2016vnl}
\bibitem{Clement:2016vnl}
H.~Clement,
``On the History of Dibaryons and their Final Observation,''
Prog. Part. Nucl. Phys. \textbf{93}, 195 (2017)
%doi:10.1016/j.ppnp.2016.12.004
%[arXiv:1610.05591 [nucl-ex]].
%88 citations counted in INSPIRE as of 31 Jul 2022



%\cite{WASA-at-COSY:2011bjg}
\bibitem{WASA-at-COSY:2011bjg}
P.~Adlarson \textit{et al.} [WASA-at-COSY],
``ABC Effect in Basic Double-Pionic Fusion --- Observation of a new resonance?,''
Phys. Rev. Lett. \textbf{106}, 242302 (2011)
%doi:10.1103/PhysRevLett.106.242302
%[arXiv:1104.0123 [nucl-ex]].
%236 citations counted in INSPIRE as of 06 Apr 2023

%\cite{Gal:2013dca}
\bibitem{Gal:2013dca}
A.~Gal and H.~Garcilazo,
``Three-Body Calculation of the Delta-Delta Dibaryon Candidate D(03) at 2.37 GeV,''
Phys. Rev. Lett. \textbf{111}, 172301 (2013)
%doi:10.1103/PhysRevLett.111.172301
%[arXiv:1308.2112 [nucl-th]].
%101 citations counted in INSPIRE as of 12 Apr 2023


%\cite{Huang:2014kja}
\bibitem{Huang:2014kja}
F.~Huang, Z.~Y.~Zhang, P.~N.~Shen and W.~L.~Wang,
``Is d* a candidate for a hexaquark-dominated exotic state?,''
Chin. Phys. C \textbf{39}, no.7, 071001 (2015)
%doi:10.1088/1674-1137/39/7/071001
%[arXiv:1408.0458 [nucl-th]].
%56 citations counted in INSPIRE as of 23 Mar 2022

%\cite{Haidenbauer:2011za}
\bibitem{Haidenbauer:2011za}
J.~Haidenbauer and U.~G.~Meissner,
``Exotic bound states of two baryons in light of chiral effective field theory,''
Nucl. Phys. A \textbf{881}, 44-61 (2012)
%doi:10.1016/j.nuclphysa.2012.01.021
%[arXiv:1111.4069 [nucl-th]].
%46 citations counted in INSPIRE as of 23 Mar 2022

%\cite{Park:2015nha}
\bibitem{Park:2015nha}
W.~Park, A.~Park and S.~H.~Lee,
``Dibaryons in a constituent quark model,''
Phys. Rev. D \textbf{92}, no.1, 014037 (2015)
%doi:10.1103/PhysRevD.92.014037
%[arXiv:1506.01123 [nucl-th]].
%44 citations counted in INSPIRE as of 23 Mar 2022

%\cite{Dong:2015cxa}
\bibitem{Dong:2015cxa}
Y.~Dong, P.~Shen, F.~Huang and Z.~Zhang,
``Theoretical study of the $d^*(2380) \to d \pi \pi$ decay width,''
Phys. Rev. C \textbf{91}, no.6, 064002 (2015)
%doi:10.1103/PhysRevC.91.064002
%[arXiv:1503.02456 [nucl-th]].
%39 citations counted in INSPIRE as of 23 Mar 2022

%\cite{Dong:2016rva}
\bibitem{Dong:2016rva}
Y.~Dong, F.~Huang, P.~Shen and Z.~Zhang,
``Decay width of $d^*(2380)\to NN \pi\pi$ processes,''
Phys. Rev. C \textbf{94}, no.1, 014003 (2016)
%doi:10.1103/PhysRevC.94.014003
%[arXiv:1603.08748 [hep-ph]].
%35 citations counted in INSPIRE as of 23 Mar 2022

%\cite{Ikeno:2021frl}
\bibitem{Ikeno:2021frl}
N.~Ikeno, R.~Molina and E.~Oset,
``Triangle singularity mechanism for the $pp\to\pi+d$ fusion reaction,''
Phys. Rev. C \textbf{104}, no.1, 014614 (2021)
%doi:10.1103/PhysRevC.104.014614
%[arXiv:2103.01712 [nucl-th]].
%11 citations counted in INSPIRE as of 29 Aug 2022

%\cite{Molina:2021bwp}
\bibitem{Molina:2021bwp}
R.~Molina, N.~Ikeno and E.~Oset,
``Sequential single pion production explaining the dibaryon ''d*(2380)'' peak*,''
Chin. Phys. C \textbf{47}, no.4, 041001 (2023)
%doi:10.1088/1674-1137/acb0b7
%[arXiv:2102.05575 [nucl-th]].
%16 citations counted in INSPIRE as of 30 Jun 2023

%\cite{Bar-Nir:1973mxc}
\bibitem{Bar-Nir:1973mxc}
I.~Bar-Nir, E.~Burkhardt, H.~Filthuth, H.~Oberlack, A.~Putzer, P.~Ang, G.~Alexander, O.~Benary, S.~Dagan and J.~Grunhaus, \textit{et al.}
``Analysis of the reaction $n p \to \pi^+ \pi^-$ below 3.5 Gev/c,''
Nucl. Phys. B \textbf{54}, 17-28 (1973)
%doi:10.1016/0550-3213(73)90062-X
%53 citations counted in INSPIRE as of 29 Aug 2022







%\cite{LHCb:2015yax}
\bibitem{LHCb:2015yax}
R.~Aaij \textit{et al.} [LHCb],
``Observation of $J/\psi p$ Resonances Consistent with Pentaquark States in $\Lambda_b^0 \to J/\psi K^- p$ Decays,''
Phys. Rev. Lett. \textbf{115}, 072001 (2015)
%doi:10.1103/PhysRevLett.115.072001
%[arXiv:1507.03414 [hep-ex]].
%1544 citations counted in INSPIRE as of 13 Apr 2023

%\cite{LHCb:2019kea}
\bibitem{LHCb:2019kea}
R.~Aaij \textit{et al.} [LHCb],
``Observation of a narrow pentaquark state, $P_c(4312)^+$, and of two-peak structure of the $P_c(4450)^+$,''
Phys. Rev. Lett. \textbf{122}, no.22, 222001 (2019)
%doi:10.1103/PhysRevLett.122.222001
%[arXiv:1904.03947 [hep-ex]].
%574 citations counted in INSPIRE as of 13 Apr 2023

%\cite{LHCb:2020jpq}
\bibitem{LHCb:2020jpq}
R.~Aaij \textit{et al.} [LHCb],
``Evidence of a $J/\psi\Lambda$ structure and observation of excited $\Xi^-$ states in the $\Xi^-_b \to J/\psi\Lambda K^-$ decay,''
Sci. Bull. \textbf{66}, 1278-1287 (2021)
%doi:10.1016/j.scib.2021.02.030
%[arXiv:2012.10380 [hep-ex]].
%141 citations counted in INSPIRE as of 13 Apr 2023

%\cite{Belle:2003nnu}
\bibitem{Belle:2003nnu}
S.~K.~Choi \textit{et al.} [Belle],
``Observation of a narrow charmonium-like state in exclusive $B^\pm \to K^\pm \pi^+ \pi^- J/\psi$ decays,''
Phys. Rev. Lett. \textbf{91}, 262001 (2003)
%doi:10.1103/PhysRevLett.91.262001
%[arXiv:hep-ex/0309032 [hep-ex]].
%2299 citations counted in INSPIRE as of 13 Apr 2023

%\cite{Tornqvist:2004qy}
\bibitem{Tornqvist:2004qy}
N.~A.~Tornqvist,
``Isospin breaking of the narrow charmonium state of Belle at 3872-MeV as a deuson,''
Phys. Lett. B \textbf{590}, 209-215 (2004)
%doi:10.1016/j.physletb.2004.03.077
%[arXiv:hep-ph/0402237 [hep-ph]].
%577 citations counted in INSPIRE as of 13 Apr 2023

%\cite{BESIII:2013ris}
\bibitem{BESIII:2013ris}
M.~Ablikim \textit{et al.} [BESIII],
``Observation of a Charged Charmoniumlike Structure in $e^+e^- \to \pi^+\pi^- J/\psi$ at $\sqrt{s}$ =4.26 GeV,''
Phys. Rev. Lett. \textbf{110}, 252001 (2013)
%doi:10.1103/PhysRevLett.110.252001
%[arXiv:1303.5949 [hep-ex]].
%1018 citations counted in INSPIRE as of 13 Apr 2023

%\cite{Xiao:2013iha}
\bibitem{Xiao:2013iha}
T.~Xiao, S.~Dobbs, A.~Tomaradze and K.~K.~Seth,
``Observation of the Charged Hadron $Z_c^{\pm}(3900)$ and Evidence for the Neutral $Z_c^0(3900)$ in $e^+e^-\to \pi\pi J/\psi$ at $\sqrt{s}=4170$ MeV,''
Phys. Lett. B \textbf{727}, 366-370 (2013)
%doi:10.1016/j.physletb.2013.10.041
%[arXiv:1304.3036 [hep-ex]].
%399 citations counted in INSPIRE as of 13 Apr 2023

%\cite{Froemel:2004ea}
\bibitem{Froemel:2004ea}
F.~Froemel, B.~Julia-Diaz and D.~O.~Riska,
``Bound states of double flavor hyperons,''
Nucl. Phys. A \textbf{750}, 337-356 (2005)
%doi:10.1016/j.nuclphysa.2005.01.022
%[arXiv:nucl-th/0410034 [nucl-th]].
%27 citations counted in INSPIRE as of 10 Apr 2023

%\cite{Liu:2011xc}
\bibitem{Liu:2011xc}
Y.~R.~Liu and M.~Oka,
``$\Lambda_c N$ bound states revisited,''
Phys. Rev. D \textbf{85}, 014015 (2012)
%doi:10.1103/PhysRevD.85.014015
%[arXiv:1103.4624 [hep-ph]].
%120 citations counted in INSPIRE as of 10 Apr 2023

%\cite{Song:2022yfr}
\bibitem{Song:2022yfr}
L.~Q.~Song, D.~Song, J.~T.~Zhu and J.~He,
``Possible $\Lambda_c\Lambda_c$ molecular states and their productions in nucleon-antinucleon collision,''
Phys. Lett. B \textbf{835}, 137586 (2022)
%doi:10.1016/j.physletb.2022.137586
%[arXiv:2207.13957 [hep-ph]].
%3 citations counted in INSPIRE as of 28 Jun 2023

%\cite{Lee:2011rka}
\bibitem{Lee:2011rka}
N.~Lee, Z.~G.~Luo, X.~L.~Chen and S.~L.~Zhu,
``Possible Deuteron-like Molecular States Composed of Heavy Baryons,''
Phys. Rev. D \textbf{84}, 014031 (2011)
%doi:10.1103/PhysRevD.84.014031
%[arXiv:1104.4257 [hep-ph]].
%59 citations counted in INSPIRE as of 06 May 2023


%\cite{Wang:2021qmn}
\bibitem{Wang:2021qmn}
X.~W.~Wang, Z.~G.~Wang and G.~l.~Yu,
``Study of $\Lambda _c\Lambda _c$ dibaryon and $\Lambda _c{\bar{\Lambda }}_c$ baryonium states via QCD sum rules,''
Eur. Phys. J. A \textbf{57}, no.9, 275 (2021)
%doi:10.1140/epja/s10050-021-00576-8
%[arXiv:2107.04751 [hep-ph]].
%8 citations counted in INSPIRE as of 29 Jul 2022


%\30cite{Chen:2011cta}
\bibitem{Chen:2011cta}
Y.~D.~Chen and C.~F.~Qiao,
``Baryonium Study in Heavy Baryon Chiral Perturbation Theory,''
Phys. Rev. D \textbf{85} (2012), 034034
%doi:10.1103/PhysRevD.85.034034
%[arXiv:1102.3487 [hep-ph]].
%16 citations counted in INSPIRE as of 19 Jun 2022


%\34cite{Simonov:2011jc}
\bibitem{Simonov:2011jc}
Y.~A.~Simonov,
``Theory of hadron decay into baryon-antibaryon final state,''
Phys. Rev. D \textbf{85} (2012), 105025
%doi:10.1103/PhysRevD.85.105025
%[arXiv:1109.5545 [hep-ph]].
%8 citations counted in INSPIRE as of 19 Jun 2022


%\cite{Li:2012bt}
\bibitem{Li:2012bt}
N.~Li and S.~L.~Zhu,
``Hadronic Molecular States Composed of Heavy Flavor Baryons,''
Phys. Rev. D \textbf{86}, 014020 (2012)
%doi:10.1103/PhysRevD.86.014020
%[arXiv:1204.3364 [hep-ph]].
%40 citations counted in INSPIRE as of 08 Aug 2022

%\cite{Garcilazo:2020acl}
\bibitem{Garcilazo:2020acl}
H.~Garcilazo and A.~Valcarce,
``Doubly charmed multibaryon systems,''
Eur. Phys. J. C \textbf{80}, no.8, 720 (2020)
%doi:10.1140/epjc/s10052-020-8320-0
%[arXiv:2008.00675 [hep-ph]].
%13 citations counted in INSPIRE as of 08 Aug 2022

%\cite{Carames:2015sya}
\bibitem{Carames:2015sya}
T.~F.~Carames and A.~Valcarce,
``Heavy flavor dibaryons,''
Phys. Rev. D \textbf{92}, no.3, 034015 (2015)
%doi:10.1103/PhysRevD.92.034015
%[arXiv:1507.08278 [hep-ph]].
%17 citations counted in INSPIRE as of 08 Aug 2022

%\cite{Chen:2021cfl}
\bibitem{Chen:2021cfl}
K.~Chen, R.~Chen, L.~Meng, B.~Wang and S.~L.~Zhu,
``Systematics of the heavy flavor hadronic molecules,''
Eur. Phys. J. C \textbf{82}, no.7, 581 (2022)
%doi:10.1140/epjc/s10052-022-10540-5
%[arXiv:2109.13057 [hep-ph]].
%22 citations counted in INSPIRE as of 08 Aug 2022

%\57cite{Ling:2021asz}
\bibitem{Ling:2021asz}
X.~Z.~Ling, M.~Z.~Liu and L.~S.~Geng,
``Masses and strong decays of open charm hexaquark states $\Sigma _{c}^{(*)}{\Sigma }_{c}^{(*)}$,''
Eur. Phys. J. C \textbf{81} (2021) no.12, 1090
%doi:10.1140/epjc/s10052-021-09867-2
%[arXiv:2110.13792 [hep-ph]].
%3 citations counted in INSPIRE as of 20 Jun 2022

%\cite{Dong:2021bvy}
\bibitem{Dong:2021bvy}
X.~K.~Dong, F.~K.~Guo and B.~S.~Zou,
``A survey of heavy\textendash{}heavy hadronic molecules,''
Commun. Theor. Phys. \textbf{73}, no.12, 125201 (2021)
%doi:10.1088/1572-9494/ac27a2
%[arXiv:2108.02673 [hep-ph]].
%59 citations counted in INSPIRE as of 09 Aug 2022
%\cite{Song:2022svi}
\bibitem{Song:2022svi}
D.~Song, L.~Q.~Song, S.~Y.~Kong and J.~He,
``Possible molecular states from interactions of charmed baryons,''
Phys. Rev. D \textbf{106}, no.7, 074030 (2022)
%doi:10.1103/PhysRevD.106.074030
%[arXiv:2208.01879 [hep-ph]].
%1 citations counted in INSPIRE as of 06 Apr 2023


%3\cite{Isola:2003fh}
\bibitem{Isola:2003fh}
  C.~Isola, M.~Ladisa, G.~Nardulli and P.~Santorelli,
  ``Charming penguins in $B\to K^* \pi, K(\rho, \omega, \phi)$ decays,''
   Phys.\ Rev.\ D {\bf 68}, 114001 (2003)  %doi:10.1103/PhysRevD.68.114001
  % [hep-ph/0307367].  %%CITATION = doi:10.1103/PhysRevD.68.114001;%%  %94 citations counted in INSPIRE as of 31 Aug 2016

%4\cite{Falk:1992cx}
\bibitem{Falk:1992cx}
  A.~F.~Falk and M.~E.~Luke,
  ``Strong decays of excited heavy mesons in chiral perturbation theory,''
  Phys.\ Lett.\ B {\bf 292}, 119 (1992)
 % doi:10.1016/0370-2693(92)90618-E
  %[hep-ph/9206241].
  %%CITATION = doi:10.1016/0370-2693(92)90618-E;%%
  %201 citations counted in INSPIRE as of 28 Mar 2019


%\cite{Zhu:2021lhd}
\bibitem{Zhu:2021lhd}
J.~T.~Zhu, L.~Q.~Song and J.~He,
``$P_{cs}(4459)$ and other possible molecular states from $\Xi_{c}^{(*)}\bar{D}^{(*)}$ and $\Xi'_c\bar{D}^{(*)}$ interactions,''
Phys. Rev. D \textbf{103}, no.7, 074007 (2021)
%doi:10.1103/PhysRevD.103.074007
%[arXiv:2101.12441 [hep-ph]].
%29 citations counted in INSPIRE as of 31 Jul 2022


  %5%\cite{Chen:2019asm}
\bibitem{Chen:2019asm}
R.~Chen, Z.~F.~Sun, X.~Liu and S.~L.~Zhu,
``Strong LHCb evidence supporting the existence of the hidden-charm molecular pentaquarks,''
Phys. Rev. D \textbf{100}, no.1, 011502 (2019)
%doi:10.1103/PhysRevD.100.011502
%[arXiv:1903.11013 [hep-ph]].
%89 citations counted in INSPIRE as of 20 Jan 2021


%\cite{Zhu:2022fyb}
\bibitem{Zhu:2022fyb}
J.~T.~Zhu, S.~Y.~Kong, L.~Q.~Song and J.~He,
``Systematical study of $\Omega_c$-like molecular states from interactions $\Xi_c^{(',*)}K^{(*)}$ and $\Xi^{(*)}D^{(*)}$,''
Phys. Rev. D \textbf{105} (2022) no.9, 094036
%doi:10.1103/PhysRevD.105.094036
%[arXiv:2205.07586 [hep-ph]].
%0 citations counted in INSPIRE as of 01 Aug 2022

%\18cite{Dong:2021juy}
\bibitem{Dong:2021juy}
X.~K.~Dong, F.~K.~Guo and B.~S.~Zou,
``A survey of heavy-antiheavy hadronic molecules,''
Progr. Phys. \textbf{41} (2021), 65-93
%doi:10.13725/j.cnki.pip.2021.02.001
%[arXiv:2101.01021 [hep-ph]].
%59 citations counted in INSPIRE as of 18 Jun 2022

%\cite{He:2019rva}
\bibitem{He:2019rva}
J.~He and D.~Y.~Chen,
``Molecular states from $\Sigma^{(*)}_c\bar{D}^{(*)}-\Lambda_c\bar{D}^{(*)}$ interaction,''
Eur. Phys. J. C \textbf{79}, no.11, 887 (2019)
%doi:10.1140/epjc/s10052-019-7419-7
%[arXiv:1909.05681 [hep-ph]].
%34 citations counted in INSPIRE as of 06 May 2023



%17\cite{Zhu:2019ibc}
\bibitem{Zhu:2019ibc}
J.~T.~Zhu, Y.~Liu, D.~Y.~Chen, L.~Jiang and J.~He,
``$X$(2239) and $\eta (2225)$ as hidden-strange molecular states from $\Lambda \bar \Lambda$ interaction,''
Chin. Phys. C \textbf{44} (2020) no.12, 123103
%doi:10.1088/1674-1137/abb4cc
%[arXiv:1911.03706 [hep-ph]].
%6 citations counted in INSPIRE as of 18 Jun 2022


  %14\cite{Phillips:1967zza}
  \bibitem{Phillips:1967zza}
  R.~J.~N.~Phillips,
  ``Antinuclear Forces,''
  Rev.\ Mod.\ Phys.\  {\bf 39}, 681 (1967).
 % doi:10.1103/RevModPhys.39.681
  %%CITATION = doi:10.1103/RevModPhys.39.681;%%
  %52 citations counted in INSPIRE as of 06 Nov 2019

   %15\cite{Klempt:2002ap}
\bibitem{Klempt:2002ap}
  E.~Klempt, F.~Bradamante, A.~Martin and J.~M.~Richard,
  ``Antinucleon nucleon interaction at low energy: Scattering and protonium,''
  Phys.\ Rept.\  {\bf 368}, 119 (2002).
%  doi:10.1016/S0370-1573(02)00144-8
  %%CITATION = doi:10.1016/S0370-1573(02)00144-8;%%
  %139 citations counted in INSPIRE as of 06 Nov 2019

%\cite{He:2014nya}
\bibitem{He:2014nya}
J.~He,
``Study of the $B\bar{B}^*/D\bar{D}^*$ bound states in a Bethe-Salpeter approach,''
Phys. Rev. D \textbf{90}, no.7, 076008 (2014)
%doi:10.1103/PhysRevD.90.076008
%[arXiv:1409.8506 [hep-ph]].
%48 citations counted in INSPIRE as of 10 Apr 2023


%\32cite{Ding:2020dio}
\bibitem{Ding:2020dio}
Z.~M.~Ding, H.~Y.~Jiang and J.~He,
``Molecular states from $D^{(*)}\bar{D}^{(*)}/B^{(*)}\bar{B}^{(*)}$ and $D^{(*)}D^{(*)}/\bar{B}^{(*)}\bar{B}^{(*)}$ interactions,''
Eur. Phys. J. C \textbf{80} (2020) no.12, 1179
%doi:10.1140/epjc/s10052-020-08754-6
%[arXiv:2011.04980 [hep-ph]].
%15 citations counted in INSPIRE as of 19 Jun 2022

%\33cite{}
\bibitem{Kong:2021ohg}
S.~Y.~Kong, J.~T.~Zhu, D.~Song and J.~He,
``Heavy-strange meson molecules and possible candidates Ds0*(2317), Ds1(2460), and X0(2900),''
Phys. Rev. D \textbf{104} (2021) no.9, 094012
%doi:10.1103/PhysRevD.104.094012
%[arXiv:2106.07272 [hep-ph]].
%9 citations counted in INSPIRE as of 19 Jun 2022







  %8\cite{He:2015yva}
\bibitem{He:2015yva}
  J.~He,
  ``Internal structures of the nucleon resonances N(1875) and N(2120),''
  Phys.\ Rev.\ C {\bf 91}, no.1,  018201 (2015)
%  doi:10.1103/PhysRevC.91.018201
  %[arXiv:1501.00522 [nucl-th]].
  %%CITATION = doi:10.1103/PhysRevC.91.018201;%%
  %18 citations counted in INSPIRE as of 09 Sep 2019

  %9\cite{He:2017aps}
\bibitem{He:2017aps}
  J.~He,
  ``Nucleon resonances $N(1875)$ and $N(2100)$ as strange partners of LHCb pentaquarks,''
  Phys.\ Rev.\ D {\bf 95}, no.7,  074031 (2017)
 % doi:10.1103/PhysRevD.95.074031
  %[arXiv:1701.03738 [hep-ph]].
  %%CITATION = doi:10.1103/PhysRevD.95.074031;%%
  %17 citations counted in INSPIRE as of 09 Sep 2019
%10\cite{He:2014nya}


 %11\cite{He:2015mja}
\bibitem{He:2015mja}
  J.~He,
  ``The $Z_c(3900)$ as a resonance from the $D\bar{D}^*$ interaction,''
  Phys.\ Rev.\ D {\bf 92}, no.3,  034004 (2015)
 % doi:10.1103/PhysRevD.92.034004
  %[arXiv:1505.05379 [hep-ph]].
  %%CITATION = doi:10.1103/PhysRevD.92.034004;%%
  %28 citations counted in INSPIRE as of 28 Mar 2019

  %12\cite{He:2012zd}
\bibitem{He:2012zd}
  J.~He, D.~Y.~Chen and X.~Liu,
  ``New Structure Around 3250 MeV in the Baryonic B Decay and the $D^*_0(2400)N$ Molecular Hadron,''
  Eur.\ Phys.\ J.\ C {\bf 72}, 2121 (2012)
 % doi:10.1140/epjc/s10052-012-2121-z
  %[arXiv:1204.6390 [hep-ph]].
  %%CITATION = doi:10.1140/epjc/s10052-012-2121-z;%%
  %20 citations counted in INSPIRE as of 09 Sep 2019

  %\cite{He:2019ify}
\bibitem{He:2019ify}
J.~He,
``Study of $P_c(4457)$, $P_c(4440)$, and $P_c(4312)$ in a quasipotential Bethe-Salpeter equation approach,''
Eur. Phys. J. C \textbf{79}, no.5, 393 (2019)
%doi:10.1140/epjc/s10052-019-6906-1
%[arXiv:1903.11872 [hep-ph]].
%132 citations counted in INSPIRE as of 07 May 2023
%%%%%%%%%%%%%%%%%%%%%%%%%%%%%%%%%%%







\end{thebibliography}
\end{document}